\documentstyle{IOPConf}    
\begin{document}
\frenchspacing

\title{Probing the Structure of Halo Nuclei}

\longauthor{I.J. Thompson$^a$\\ and the Russian-Nordic-British Theory (RNBT) collaboration}{I.J. Thompson}

\address{$^a$Department of Physics, University of Surrey, Guildford GU2 5XH, United Kingdom}


\beginabstract
Our understanding of halo nuclei has so far relied on high-energy scattering and
reactions, but a number of uncertainties remain.  I discuss in general terms
the new range of observables which will be measured by experiments around the
Coulomb barrier, and how some details of the reaction mechanisms still need to
be clarified.
\endabstract

Halo nuclei are nuclei in which there is very weak binding of the last 
nucleon or pair of nucleons. There is then a large probability for these
nucleons tunneling into the classically-forbidden regions outside a more
tightly bound core. The overall r.m.s. matter radius of the nucleus
is then very large, resulting in the large interaction cross sections
which were the first experimental signatures of such nuclei \cite{tan88},
and which continue to yield important information \cite{jak96}.

Since then, two further experimental probes of halo structure have been
developed, still using high-energy reactions. 
The first is to look at the momentum distributions following
fragmentation of the halo nucleus at high velocity. In such reactions
the sudden or Serber model for breakup implies that the fragments move
after the collision with velocities reflecting their range of Fermi
momenta in the initial halo nucleus. Experimentally,  very narrow
momentum widths are found \cite{kob88,orr92}.  By the Heisenberg
Uncertainty Principle, a narrow momentum uncertainty reflects a large
spatial extension.

The second probe is the Coulomb breakup cross section when
the nuclei are incident on highly charged targets. These measurements
use a heavy target as a source of virtual photons, and reconstruct the
dipole strength function by measuring the angles and velocities of all
the fragments. Experiments find a strong concentration of dipole breakup
strength in low continuum energies. This is because the halo neutrons,
although not charged themselves, are sufficiently far from the charged
core nucleus, that the centres of mass and charge no longer coincide.
The repulsion of the target on the core alone is then
sufficient to break up the halo projectile. There has been considerable
debate among theorists about whether or not, having broken up, the particles
still attract each other sufficiently to form a `soft dipole' excited
state or resonance at low energies in the breakup continuum.

All these features are most clearly seen for the last two neutrons in
$^6$He and $^{11}$Li \cite{phyrep}, and for the last neutron in
$^{11}$Be.  They are accentuated when the halo nucleons can occupy
$s$-states relative to the central core. We have recently investigated
$^{14}$Be, where possible low-lying $s$-wave states in $^{13}$Be
contribute distinctively \cite{beryl}, and also studied \cite{neon} the
proton drip-line nucleus $^{17}$Ne in a model of $^{15}$O plus two
protons.

In the three-body cases (all except  $^{11}$Be) there are two nucleons
outside a core, bound in such a way that if one of the bodies is
removed, the other pair no longer forms a bound state. This we call
the `Borromean' configuration after the Borromean rings of tradition
and  knot theory. This configuration means that at least three-body
correlations must be treated explicitly. We find that three-body
models \cite{phyrep} are much more successful than models in which the
nucleons move independently in a mean field. In three body theories,
the bound and continuum states can be treated with proper consideration
of the transition at threshold.

\subsubsection*{Current Uncertainties}

Even taking into account the above experiments, a number of uncertainties 
remain concerning the detailed structure of halo nuclei:
\begin{enumerate}
\item
The admixture of intruder levels in the ground state of two-nucleon halo
nuclei is often difficult to precisely determine, because usually the
core + one-nucleon system is unbound (the Borromean configuration).
\item
The details of the pairing correlations are not always clear.
Three-body models have tended to use either free nucleon-nucleon
potentials, but density-dependent effective interactions appropriate to nuclear matter
have also been used.
\item
The `core' in $^6$He is the $\alpha$-particle, which is relatively
inert, but still the binding energy of $^6$He has important
contributions from $t+t$ degrees of freedom \cite{csoto93,hoff97}.  The cores in
the heavier halo nuclei are softer \cite{corex}, but it is not clear to
what extent they are excited  by the halo nucleons.
\item
In any reaction of a halo nucleus, excited halo states will typically
be produced. Since, however, halo nuclei typically have only one bound
state, all excited states are in the continuum. The physical role of
continuum intermediate and final states needs to be clarified, especially the
role of the two-nucleon continuum.
\item
The calculation of E1 dipole breakup of two-neutron halo nuclei requires the
calculation of the continuum states of the three-body n+n+core system. Because
of the large size of the ground state, the wavefunctions in the continuum
region $0 < E < 3$ MeV must be accurately calculated, and it is precisely in
this region that the existence of any soft-dipole mode must be examined. 
We find \cite{h3cont} that the scattering in this region is strongly
influenced by the neutron-neutron correlation, but that it does not
appear that this correlation is sufficiently strong to constitute a
soft dipole resonance. Other continuum resonances {\em are} predicted
by three-body continuum models, with a wide variety of structures. 
\end{enumerate}

\subsubsection*{Low-energy reactions}

When low-energy beams of halo nuclei become available, several new
kinds of experiments become possible. In particular, reaction studies at
incident energies near the Coulomb barrier. More precise elastic and
inelastic angular distributions can be measured, along with fusion
probabilities and transfer cross sections.  Let us consider how such
measurements may be used to resolve some of the uncertainties listed
above.

\subsubsection*{Halo elastic scattering}

The elastic scattering angular distribution for energies around the Coulomb
barrier is sensitive to the nuclear attraction at the surface, 
as well as to the depletion caused by any long-range excitation mechanisms.
Halo nuclei have a more diffuse density, so even the Watanabe (folded)
potential should have a diffuse real part \cite{jim-fold}. There has
already been a problem seen \cite{kolata}
in the forward-angle scattering of $^{11}$Li
on $^{12}$C which may \cite{merm93,coop93} or may not \cite{jim93} be related
to this diffuseness.

Depletion at forward angles in elastic scattering is caused by any
long-range reaction channels, so, for highly charged targets,
there should be large effects of this kind
\cite{quino} caused by the E1 couplings to and from low-lying breakup
channels in the continuum.
The elastic scattering cross section is
reduced down to quite forward angles, because the E1 excitation
mechanism is of such a long range. A similar depletion effect occurs in the
dipole excitation of the first excited state of $^{11}$Be \cite{berex}.
Of course this will only be seen if the 320 keV separation of
the excited state can be resolved experimentally. It would be of
interest to confirm the procedure with $^{11}$Be scattering, and then
to perform this experiment for the elastic scattering of $^{11}$Li, 
where different direct
measurements of the E1 distribution \cite{ieki93,riken94} produce
disparate results.

\subsubsection*{Fusion cross sections around the barrier}

There has been considerable debate in the literature concerning the
possible enhancement (and/or reduction) of fusion cross sections for
halo nuclei at barrier and sub-barrier energies. It is well known that
couplings to inelastic states lead to a reduction in the effective
barrier, and an enhancement of the fusion cross section. Some
theoretical work \cite{taki93,dasso94,dasso96} holds that similar
considerations apply during the barrier traversal of halo nuclei,
whereas others \cite{huss92,canto95} come to another conclusion. The
latter believe that the large probability of breakup to the dipole
channels (mentioned in the previous paragraph) depletes the elastic 
channel, and reduces fusion.

To date, the experimental evidence is unclear. 
Measurements \cite{sida95} of the fusion of $^{11}$Be on $^{238}$U, are
of insufficient accuracy to determine whether there is a barrier
enhancement. RIKEN experimenters \cite{petra96} have attempted to
determine $^9$Li and $^{11}$Li fusion on a $^{28}$Si target, and found
similar fusion probabilities for the two projectiles. Recently, fusion
cross sections for $^{6,7}$Li + $^9$Be and for $^{6,7}$Li + $^{12}$C
have been systematically measured and analysed to look for the
dependence on the relevant separation energies \cite{taka97}. They
found  reductions of fusion up to 40\%, for energies from near the barrier
to approximately twice the  barrier.

The modelling of the doorway processes leading to the fusion of halo
nuclei is a non-trivial problem. The various models used
\cite{taki93,dasso94,dasso96,taka97} make different assumptions about the
lifetime of the excited states produced when the halo nucleons are
perturbed in the reaction. These states are breakup states, it should
be noted, and demand, ideally at least, a three-body model which takes
elastic scattering, breakup and fusion processes into account in a
unified manner. Such a model would describe the role of intermediate 
continuum states, even when these states are non-resonant, and do not
even have well-defined decay widths.
Such a model would moreover describe the reversible (virtual)
production of excited states at lower energies (in the Born-Oppenheimer
limit, these states would be produced {\em completely} reversibly), the
irreversible (real) breakup at higher beam energies, and the smooth
transition between these limits at energies of interest. Unfortunately,
such a model is not yet available; the best we have are
coupled-reaction-channels models \cite{iman95} which discretise the
continuum in a CDCC manner.  These CDCC models, however, only include some of the
continuum outgoing channels (when the projectile c.m. leaves the
target, not when the fragments leave individually).

The fusion of two-nucleon halo nuclei such as $^6$He and $^{11}$Li
would furthermore demand a {\em four}-body reaction model. We have
four-body models for high-energy reactions \cite{att}, but these make
the other (sudden) adiabatic approximation. A theory of low-energy 
reactions of halo nuclei would have to take into account the details
of pairing in the initial and final states, as well as the mechanisms
of both simultaneous and sequential pair transfers. These mechanisms
have not been properly resolved even for `normal' nuclei, and are complicated
for Borromean halo nuclei by the lack of discrete intermediate channels
during sequential transfers.

\subsubsection*{Transfer reactions}

We have previously studied the phenomenon of $s$-wave intruder orbits
in the structure of $^{11}$Li \cite{intr}, the best-known halo
nucleus.  Recently we have developed similar models for the $^{12}$Be
and $^{14}$Be isotopes \cite{beryl}.  Although $^{11}$Be has a
predominantly $s$-wave ground state, a three-body model of two neutrons
plus an {\em inert} $^{10}$Be core can only reproduce the properties of
$^{12}$Be if the valence neutrons occupy mainly the  $(p_{1/2})^2$
configuration with about 25$\%$ admixture of $(sd)^2$ configurations
\cite{beryl}.

Core excitation can also be included \cite{corex} when solving the
$^{12}$Be three body problem since it has a large contribution to the
g.s. of the subsystem $^{11}$Be \cite{core1}.  Including the $2^+$
first excited state of $^{10}$Be in the calculation, we find
\cite{corex} that a significant part of the $^{12}$Be g.s. wave
function has core excited components ($\sim$ 40\%) and that the valence
neutrons are mainly in $(sd)^2$ configurations with only 10\% of
$(p_{1/2})^2$.  The neutron transfer reaction ($^{12}$Be,$^{11}$Be)
would \cite{corex} discriminate between the inert-core and core-coupled
models, even if the discrete states in $^{11}$Be cannot be separately
resolved. Proton targets could be used for the (p,d) neutron transfer,
enabling us to extract spectroscopic information without structure
ambiguities associated with the light particle vertex.

\subsubsection*{Inelastic cross sections}

The measurement of the cross sections for stripping to core-excited
states will also help to resolve some of the differences between the
models for $^{12}$Be discussed above. It will be possible to determine,
for example, the fraction of core excitation within the $^{12}$Be
ground state.

The models of $^{12}$Be also differ in their B(E2) transition matrix
elements between the ground state and the first $2^+$ excited state. At
high energies, however, the Coulomb B(E2) value cannot be unambiguously
determined by experimental $2^+$ angular distributions because of
extensive Coulomb-nuclear interference effects. At lower  energies
nearer the barrier, a cleaner measurement should be possible.

\subsubsection*{Summary}

The ability to perform experiments with low-energy halo nuclei,  to
look at elastic scattering and individual reaction channels, will lead
to a new range of tools for investigating their single-particle and
particle-pair spectroscopic structures. For some reactions, such as
transfers, existing reaction theories can be used to obtain accurate
results; for other reactions, such as pair transfers and fusion, more
complete reaction models still await development and application.

\ack
I thank the FUSION97 conference organisers for the unexpected
opportunity to discuss these issues.  UK support from the EPSRC grants
GR/J/95867, GR/K/95734 and GR/K33026 is  acknowledged.
\references
 \vspace{-5mm}

\list
 {[\arabic{enumi}]}{\settowidth\labelwidth{[99]}\leftmargin\labelwidth
 \advance\leftmargin\labelsep
 \usecounter{enumi}}
 \def\newblock{\hskip .11em plus .33em minus .07em}
 \sloppy\clubpenalty4000\widowpenalty4000
 \sfcode`\.=1000\relax
 \let\endthebibliography=\endlist

\itemsep=-1pt

\bibitem{tan88} 
I. Tanihata {\em et al}, Phys. Lett. {\bf B206} (1988) 592
\bibitem{jak96} J.S. Al-Khalili and J.A. Tostevin, Phys. Rev. Letts {\bf 76}  (1996) 3903
\bibitem{kob88} 
T. Kobayashi {\em et al}, Phys. Rev. Lett. {\bf 60} (1988) 2599
\bibitem{orr92} 
N.A. Orr {\em et al}, Phys. Rev. Lett. {\bf 69} (1992) 2050
\bibitem{phyrep}
 M.V. Zhukov, B.V. Danilin, D.V. Fedorov, J.M. Bang,
 I.J. Thompson, J.S. Vaagen, Phys.\ Rep.\ {\bf 231}, 151 (1993)
\bibitem{beryl} 
I.J. Thompson and M.V. Zhukov, Phys. Rev. {\bf C53} (1996) 708
\bibitem{neon} 
M.V. Zhukov and I.J. Thompson, Phys. Rev. {\bf C52} (1995) 3505
\bibitem{csoto93}
A. Cs\'ot\'o, Phys. Rev. {\bf C48} (1993) 165
\bibitem{hoff97}
J. Wurzer and H.N. Hofmann, Phys. Rev. {\bf C55} (1997) 688
\bibitem{corex}
F.M. Nunes, J.A. Christley, I.J. Thompson, R.C. Johnson and V.D. Efros,
Nucl. Phys, {\bf A609} (1996) 43
\bibitem{h3cont} 
B.V. Danilin, I.J. Thompson, M.V. Zhukov and J.S. Vaagen, to be submitted.
\bibitem{jim-fold} 
J.S. Al-Khalili, Nucl. Phys. {\bf A 581} (1995) 315
\bibitem{kolata}
J.J. Kolata {\em et al}, Phys. Rev. Lett. {\bf 69} (1992) 2631
\bibitem{merm93} M.C. Mermaz, Phys. Rev. C {\bf 47}, 2214 (1993).
\bibitem{coop93} S.G. Cooper and R.S. Mackintosh, Nucl. Phys. {\bf A582} (1995) 283
\bibitem{jim93} J.S. Al-Khalili and J.A. Tostevin, Phys. Rev.
 {\bf C49} (1993) 386
\bibitem{quino}
M.V. Andr\'es J. G\'omez-Camacho, M. A. Nagarajan, 
Nucl. Phys. {\bf A579} (1994) 273;
M.V. Andr\'es, J. G\'omez-Camacho, M. A. Nagarajan, Nucl. Phys.
{\bf A583} (1995) 817c
\bibitem{berex}
M.V. Andr\'es, J.A. Christley, J. G\'omez-Camacho, M. A. Nagarajan, Nucl. Phys.
{\bf A612} (1997) 82;
M.V. Andr\'es, J.A. Christley, J. G\'omez-Camacho, M.A. Nagarajan, 
Nucl. Phys. A (1997) in press
\bibitem{ieki93}
 K. Ieki {\it et al}, Phys.\ Rev.\ Lett.\ {\bf 70} (1993) 730;
 D. Sackett {\it et al}, Phys.\ Rev.\ C {\bf 48} (1993) 118.
\bibitem{riken94} F. Shimoura {\it et al.} Phys.\ Lett.\ B {\bf 348}
 (1995) 29.
\bibitem{taki93}
N. Takigawa, M. Kuratani and H. Sagawa, Phys. Rev. {\bf C47} (1993) R2470
\bibitem{dasso94}
C.H. Dasso and A. Vitturi, Phys. Rev. {\bf C50} (1994) R12
\bibitem{dasso96}
C.H. Dasso, J.L. Guisado, S.M. Lenzi and A. Vitturi, Nucl. Phys. {\bf A597}
(1996) 473
\bibitem{huss92}
M.S. Hussein, M.P. Pato, L.F. Canto and R. Donangelo,
Phys. Rev. {\bf C46} (1992) 377
\bibitem{canto95}
L.F. Canto, R. Donangelo, P. Lotti and M.S. Hussein, Phys. Rev. {\bf C52} (1995) R2848
\bibitem{sida95}
V. Fekou-Youmbi, J.L. Sida {\em et al}, Nucl. Phys. {\bf A583} (1995) 811
\bibitem{petra96}
M. Petrascu {\em et al}, RIKEN preprint AF-NP-237 (October 1996)
\bibitem{taka97}
J. Takahashi {\em et al}, Phys. Rev. Letts. {\bf 78} (1997) 30
\bibitem{iman95}
B. Imanishi and W. von Oertzen, Phys. Rev. {\bf C52} (1995) 3249
\bibitem{att}
J.S. Al-Khalili, I.J. Thompson and J.A. Tostevin, Nucl. Phys. {\bf A581}, (1995) 331
\bibitem{intr}
I.J. Thompson and M.V. Zhukov, Phys. Rev. {\bf C49}  (1994) 1904
\bibitem{core1}
F.M. Nunes, I.J. Thompson and R.C. Johnson, Nucl. Phys. {\bf A596} (1996) 171

\endlist

\end{document}